\documentclass[12pt,a4paper]{article}
\usepackage{amsfonts}
\usepackage{amssymb}
\usepackage{amsmath}
\usepackage{latexsym}
\begin{document}

\begin{center}
{\Large \bf Bulk metric of brane world models and\\ \vspace{2mm}
submanifolds in 6D pseudo-Euclidean space-time} \\

\vspace{4mm}

Mikhail N.~Smolyakov\\
\vspace{0.5cm} Skobeltsyn Institute of Nuclear Physics, Moscow
State University,
\\ 119991, Moscow, Russia\\
\end{center}

\begin{abstract}
In this short note, five-dimensional brane world models with
$dS_{4}$ metric on the branes are discussed. The explicit
coordinate transformations, which show the equivalence between the
bulk metric of these brane world models and the metric induced on
an appropriate submanifolds in the flat six-dimensional
pseudo-Euclidean space-time, are presented. The cases of the zero
and non-zero cosmological constant in the bulk are discussed in
detail.
\end{abstract}

It is well known that the space-time of the Randall-Sundrum model
\cite{Randall:1999ee} is a slice of the maximally-symmetric
$AdS_{5}$ space-time. It means that the bulk metric of this model
corresponds to the metric induced on a hyperboloid embedded in a
six-dimensional pseudo-Euclidean space-time. Moreover, it was
shown in \cite{Smolyakov} that the bulk metric of the form
\begin{equation}\label{metric-brane1}
ds^{2}=A^{2}(y)\left(-dt^2+\eta_{ij}dx^{i}dx^{j}\right)+dy^{2},
\end{equation}
where $\eta_{ij}=diag(1,1,1)$, $i,j=1,2,3$, can be represented as
the metric induced on the hypersurface
\begin{equation}\label{manifold-brane}
T^2-\eta_{ij}X^{i}X^{j}-Y^{2}+Z^{2}=\left[-A(y)\int\frac{dy}{dA/dy}\right]_{y=A^{-1}\left(\frac{2(Z-Y)}{\alpha}\right)},
\end{equation}
where $\alpha\ne 0$ is an arbitrary constant, in six-dimensional
pseudo-Euclidean space-time with the metric
\begin{equation}\label{metric-brane}
ds^{2}=-dT^2+\eta_{ij}dX^{i}dX^{j}+dY^{2}-dZ^{2}
\end{equation}
provided that
\begin{eqnarray}
Y&=&\frac{1}{\alpha}\left(A(y)\left(-t^2+\eta_{ij}x^{i}x^{j}\right)-\int\frac{dy}{dA/dy}\right)-\alpha\frac{A(y)}{4},\\
Z&=&\frac{1}{\alpha}\left(A(y)\left(-t^2+\eta_{ij}x^{i}x^{j}\right)-\int\frac{dy}{dA/dy}\right)+\alpha\frac{A(y)}{4},\\
T&=&A(y)t,\\
X^{i}&=&A(y)x^{i}.
\end{eqnarray}
For the Randall-Sundrum model with $A=e^{-ky}$ (for simplicity we
use $y$ instead of $|y|$) we obtain the well-known result
\begin{equation}\label{manifold-RS}
T^2+Z^{2}-\eta_{ij}X^{i}X^{j}-Y^{2}=\frac{1}{k^{2}}.
\end{equation}

Meanwhile there is a large class of brane world models with
$dS_{4}$ metric on the branes, some of them can be found in
\cite{Kaloper,ds4}. The bulk metric of such models has the form
\begin{equation}\label{bulkds}
ds^2=A^2(y)\left(-dt^2+e^{2\lambda
t}\eta_{ij}dx^{i}dx^{j}\right)+dy^2
\end{equation}
and it seems to be interesting to find an analogous hypersurface
for such form of the metric.

To solve this problem first let us consider a much simpler case --
the case of the flat five-dimensional pseudo-Euclidean space-time
with the metric
\begin{equation}\label{M}
ds_{(5)}^2=-dT^2+\eta_{ij}dX^{i}dX^{j}+dY^2.
\end{equation}
With the help of coordinate transformations
\begin{eqnarray}\label{M01}
T&=&\alpha\lambda y\left(\eta_{ij}x^{i}x^{j}e^{\lambda
t}-\frac{e^{-\lambda
t}}{\lambda^2}\right)+y\frac{\lambda}{4\alpha}e^{\lambda t},\\
\label{M02} Y&=&\alpha\lambda y\left(\eta_{ij}x^{i}x^{j}e^{\lambda
t}-\frac{e^{-\lambda
t}}{\lambda^2}\right)-y\frac{\lambda}{4\alpha}e^{\lambda t},\\
\label{M03} X^{i}&=&\lambda y x^{i}e^{\lambda t},
\end{eqnarray}
where $\alpha\ne 0$ is a constant, we can get from (\ref{M})
\begin{equation}\label{M1}
ds^2=\lambda^2y^2\left(-dt^2+e^{2\lambda
t}\eta_{ij}dx^{i}dx^{j}\right)+dy^2
\end{equation}
(note that metric (\ref{M1}) is very similar to the metric of the
well-known Milne universe \cite{Milne}, see the Appendix).

An important point is that metric (\ref{M1}) in the bulk admits
the existence of two branes with negative and positive tensions
(and with different absolute values of the tensions), which can be
located at the points $0<y_{1}<y_{2}$ of the orbifold with
$y_{1}\le y \le y_{2}$ for $y>0$ and points $-y$ and $y$
identified. Indeed, it follows from the Einstein equations (see,
for example, \cite{Kaloper}) that the brane tensions should be
such that
$\epsilon_{1}\sim-\frac{A'}{A}|_{y=y_{1}}=-\frac{1}{y_{1}}$ and
$\epsilon_{2}\sim\frac{A'}{A}|_{y=y_{2}}=\frac{1}{y_{2}}$. Thus,
even the empty bulk supports the existence of branes, moreover,
the branes appear to be inflating in this case (see examples of
such brane world models with empty five-dimensional bulk and
discussion of their properties in \cite{emptyb}).

Now with the help of formulas (\ref{M01})-(\ref{M03}) one can
easily obtain the equation for the submanifold which corresponds
to metric (\ref{bulkds}) (below we will not take into account the
existence of branes and concentrate on the metric in the bulk).
Indeed, let us consider six-dimensional pseudo-Euclidean
space-time with the metric
\begin{equation}\label{56D}
ds_{(6)}^2=-dT^2+\eta_{ij}dX^{i}dX^{j}+dY^2\mp dZ^2
\end{equation}
and the coordinate transformations
\begin{eqnarray}\label{M001}
T&=&\alpha\lambda F(y)\left(\eta_{ij}x^{i}x^{j}e^{\lambda
t}-\frac{e^{-\lambda
t}}{\lambda^2}\right)+F(y)\frac{\lambda}{4\alpha}e^{\lambda t},\\
\label{M002} Y&=&\alpha\lambda
F(y)\left(\eta_{ij}x^{i}x^{j}e^{\lambda t}-\frac{e^{-\lambda
t}}{\lambda^2}\right)-F(y)\frac{\lambda}{4\alpha}e^{\lambda
t},\\ \label{M003} X^{i}&=&\lambda F(y) x^{i}e^{\lambda t},\\
\label{M004} Z&=&G(y).
\end{eqnarray}
One can see that (\ref{M001})-(\ref{M003}) are simply
(\ref{M01})-(\ref{M03}) with $y\to F(y)$. Substituting
(\ref{M001})-(\ref{M004}) into (\ref{56D}) we get
\begin{equation}\label{M111}
ds^2=\lambda^2F^{2}(y)\left(-dt^2+e^{2\lambda
t}\eta_{ij}dx^{i}dx^{j}\right)+\left(F'^2(y)\mp
G'^2(y)\right)dy^2,
\end{equation}
where $'=\frac{d}{dy}$. So with
\begin{equation}\label{Fy}
F(y)=\frac{A(y)}{\lambda}
\end{equation}
and
\begin{equation}\label{eqFG}
F'^2\mp G'^2=1
\end{equation}
we get (\ref{bulkds}). If $F'\ge 1$, then one should take
time-like extra dimension with the coordinate $Z$ in (\ref{56D})
and $F'^2-G'^2=1$; if $F'\le 1$, then one should take space-like
extra dimension with the coordinate $Z$ in (\ref{56D}) and
$F'^2+G'^2=1$. The equation of the submanifold, which corresponds
to (\ref{M001})-(\ref{M004}), can be easily obtained and has the
form
\begin{equation}\label{subm6D}
T^2-Y^2-\eta_{ij}X^{i}X^{j}=-F^2(y)\left|_{y=G^{-1}(Z)}\right.,
\end{equation}
where $F(y)$ and $G(y)$ satisfy equations (\ref{Fy}) and
(\ref{eqFG}). This submanifold is embedded into (4+2) or (5+1)
pseudo-Euclidean space-time depending on the form of $A(y)$.

Now let us turn to specific examples. First, we consider the brane
world model, which corresponds to the negative value of the bulk
cosmological constant \cite{Kaloper}, but, contrary to the case of
the Randall-Sundrum model \cite{Randall:1999ee}, provides $dS_{4}$
metric on the branes. In this model the metric in the bulk has the
form
\begin{equation}\label{K}
ds^2=\frac{\lambda^2}{k^2}\sinh^2(ky)\left(-dt^2+e^{2\lambda
t}\eta_{ij}dx^{i}dx^{j}\right)+dy^2
\end{equation}
(note that metric (\ref{K}) has a simpler form than that of
\cite{Kaloper}, but the metric of \cite{Kaloper} can be brought to
form (\ref{K}) by a simple redefinition of the parameters and a
shift of the extra dimension coordinate). The value of the
four-dimensional Hubble parameter $\lambda$ is defined by the
boundary conditions on the branes, which are not presented here.
From (\ref{Fy}) we get $ F(y)=\frac{\sinh(ky)}{k}.$ Thus, we
should take time-like extra dimension $Z$ and
$G(y)=\frac{\cosh(ky)}{k}$. Then we easily get from (\ref{subm6D})
hyperboloid (\ref{manifold-RS}) embedded in (4+2) pseudo-Euclidean
space-time. This is the expected result because metric (\ref{K})
corresponds to the same matter in the bulk as the metric of the
Randall-Sundrum model
\begin{equation}\label{metrRS}
ds^2=e^{-2k\tilde
y}\left(-dt^2+\eta_{ij}dx^{i}dx^{j}\right)+d\tilde y^2.
\end{equation}
Indeed, the direct coordinate transformations between metric
(\ref{metrRS}) and metric (\ref{K}) are
\begin{eqnarray}\label{RSK}
\tilde
y&=&-\frac{1}{k}\ln\left(\pm\frac{\lambda}{k}\sinh(ky)e^{\lambda
t}\right),\\ \label{RSK1} \tilde t&=&\pm\frac{e^{-\lambda
t}}{\lambda}\coth(ky).
\end{eqnarray}
One can check that substitution of coordinate transformations
(\ref{RSK}), (\ref{RSK1}) into metric (\ref{metrRS}) leads to
(\ref{K}). But note that although solutions (\ref{metrRS}) and
(\ref{K}) can be transformed one to another in the bulk, the
models of \cite{Randall:1999ee} and \cite{Kaloper} correspond to
different physical systems because of the different relations
between the values of the brane tensions used in these models.

Now we turn to the second model, which was also discussed in
\cite{Kaloper}. It describes a slice of the maximally-symmetric
$dS_{5}$ space-time (because of the positive value of the
cosmological constant in the bulk) and also provides $dS_{4}$
metric on the branes. The metric of the model has the form
\begin{equation}\label{K2}
ds^2=\frac{\lambda^2}{k^2}\sin^2(ky)\left(-dt^2+e^{2\lambda
t}\eta_{ij}dx^{i}dx^{j}\right)+dy^2.
\end{equation}
((\ref{K2}) also has a simpler form than that of \cite{Kaloper},
but again up to a simple redefinition of the parameters and a
shift of the extra dimension's coordinate). Now we should take
$F(y)=\frac{\sin(ky)}{k},$ space-like extra dimension $Z$ and
$G(y)=\frac{\cos(ky)}{k}$. Then we get from (\ref{subm6D}) the
hyperboloid
\begin{equation}\label{submDs}
T^2-Z^2-\eta_{ij}X^{i}X^{j}-Y^2=-\frac{1}{k^2},
\end{equation}
embedded in (5+1) pseudo-Euclidean space-time. Of course, this is
also the expected result.

We hope that the results presented in this note can be useful from
theoretical and pedagogical points of view.

\section*{Acknowledgments}

The author is grateful to I.P.~Volobuev for discussions. The work
was supported by grant of Russian Ministry of Education and
Science NS-4142.2010.2, RFBR grant 08-02-92499-CNRSL$\_$a and
state contract 02.740.11.0244.

\section*{Appendix: Milne universe}

Metric of the Milne universe has the form
\begin{equation}\label{MU}
ds^2=-dt^2+\lambda^2t^2\left((dx^{1})^2+e^{2\lambda
x^{1}}\left((dx^{2})^2+(dx^{3})^2\right)\right).
\end{equation}
With the help of transformations
\begin{eqnarray}
x^{1}&=&\frac{1}{\lambda}\ln\left(\frac{2\alpha}{\lambda}\left(\cosh(\chi)-\sinh(\chi)\cos(\theta)\right)\right),\\
x^{2}&=&\frac{1}{2\alpha}\left(\frac{\sinh(\chi)\sin(\theta)\cos(\varphi)}{\cosh(\chi)-\sinh(\chi)\cos(\theta)}\right),\\
x^{3}&=&\frac{1}{2\alpha}\left(\frac{\sinh(\chi)\sin(\theta)\sin(\varphi)}{\cosh(\chi)-\sinh(\chi)\cos(\theta)}\right),
\end{eqnarray}
where $\alpha$ is a constant, metric (\ref{MU}) can be brought to
the more familiar form (see, for example, "expanding Minkowski
universe" in \cite{Misner})
\begin{equation}
ds^2=-dt^2+t^2\left(d\chi^2+\sinh^2(\chi)\left(d\theta^2+\sin^2(\theta)d\varphi^2\right)\right).
\end{equation}


\begin{thebibliography}{0}
\bibitem{Randall:1999ee}
L. Randall and R. Sundrum, {\it Phys. Rev. Lett.} {\bf 83} (1999)
3370.

\bibitem{Smolyakov}
M.N. Smolyakov, {\it Class. Quant. Grav.} {\bf 25} (2008) 238003;
Erratum-ibid. {\bf 27} (2010) 109801.

\bibitem{Kaloper}
N. Kaloper, {\it Phys. Rev. D} {\bf 60} (1999) 123506.

\bibitem{ds4}
A. Karch and L. Randall, {\it JHEP} {\bf 0105} (2001) 008.\\
J.M. Cline and H. Firouzjahi, {\it Phys. Lett. B} {\bf
514} (2001) 205.\\
P. Kanti, S.c. Lee and K.A. Olive, {\it Phys. Rev. D} {\bf 67}
(2003) 024037.

\bibitem{Milne}
E.A. Milne, {\it Nature} {\bf 130} (1932) 9.

\bibitem{emptyb}
P. Binetruy, C. Deffayet and D. Langlois, {\it Nucl. Phys. B} {\bf
565} (2000) 269.\\
N. Deruelle and T. Dolezel, {\it Phys. Rev. D} {\bf 62} (2000)
103502.

\bibitem{Misner}
C.W. Misner, K.S. Thorne and J.A. Wheeler, {\it Gravitation} (W.H.
Freeman and Company, San Francisco, 1973), pp. 743-744.
\end{thebibliography}
\end{document}